\documentclass[conference]{IEEEtran}
\usepackage{ifpdf}

\usepackage{cite}

\ifCLASSINFOpdf
   \usepackage[pdftex]{graphicx}
\else
   \usepackage[dvips]{graphicx}
\fi
\usepackage{amsmath}
\begin{document}

\title{Enabling Adaptive and Enhanced Acoustic Sensing \\ Using Nonlinear Dynamics}

 \author{\IEEEauthorblockN{Claudia Lenk, Lars Seeber, and Martin Ziegler}
 \IEEEauthorblockA{Dept. of Micro- and Nanoelectronic Systems\\
 Technische Universit\"at Ilmenau\\
Ilmenau, Germany\\
 Email: claudia.lenk@tu-ilmenau.de}
 \and
 \IEEEauthorblockN{Philipp H\"ovel}
 \IEEEauthorblockA{School of Mathematical Sciences\\
 University College Cork\\
 Cork, Ireland}
 \and
 \IEEEauthorblockN{Stefanie Gutschmidt}
 \IEEEauthorblockA{Dept. of Mechanical Engineering\\
 University of Canterbury\\
 Christchurch, New Zealand\\
}}

%

\IEEEoverridecommandlockouts
\IEEEpubid{\makebox[\columnwidth]{\copyright2020 IEEE \hfill} \hspace{\columnsep}\makebox[\columnwidth]{ }}
\maketitle
\IEEEpubidadjcol
\begin{abstract}
Transmission of real-time data is strongly increasing due to remote processing of sensor data, among other things. A route to meet this demand is adaptive sensing, in which sensors acquire only relevant information using pre-processing at sensor level. We present here adaptive acoustic sensors based on mechanical oscillators with integrated sensing and actuation. Their dynamics are shifted into a nonlinear regime using feedback or coupling. This enhances dynamic range, frequency resolution and signal-to-noise ratio. Combining tunable sensing properties with sound analysis could enable acquiring of only relevant information rather than extracting this from irrelevant data by post-processing.   

\end{abstract}


%
\IEEEpeerreviewmaketitle

\section{Introduction}
Human sound detection covers large dynamic ranges regarding sound pressure levels (0-120dB SPL) and frequency range (20Hz-20kHz) and simultaneously provides high-resolution detection, i.e. frequency differences of 3-5Hz and differences in sound pressure level of only 1dB can be resolved. These remarkable properties are (amongst other things) a result of the nonlinear dynamics of the sound sensors: the hair cells \cite{Ashmore}. Their dynamics are tuned by diverse coupling mechanisms such as base coupling due to attachment to basilar and tectorial membrane and hydrodynamic coupling as well as feedback mechanisms such as efferent feedback \cite{Fettiplace}. The dynamics of hair cells and the resulting sensing properties resemble thereby the dynamics and properties of a critical oscillator tuned near or at a Hopf bifurcation \cite{Juelicher, Hopf}. These dynamics yield a nonlinear amplification \cite{Robles} of sound signals during the sensing process itself. In contrast, in conventional sound detection with microphones and amplifiers amplification is applied to the detected signal in subsequent processing steps. By integrating the amplification into the sensing process, detrimental signal distortions can be prevented, which may arise due to filtering and amplifying the signal in processing steps. The compressive amplification enhances the detection of signals with low signal-to-noise ratios, in particular.

\begin{figure}[!t]
\centering
\includegraphics[width=3.5in]{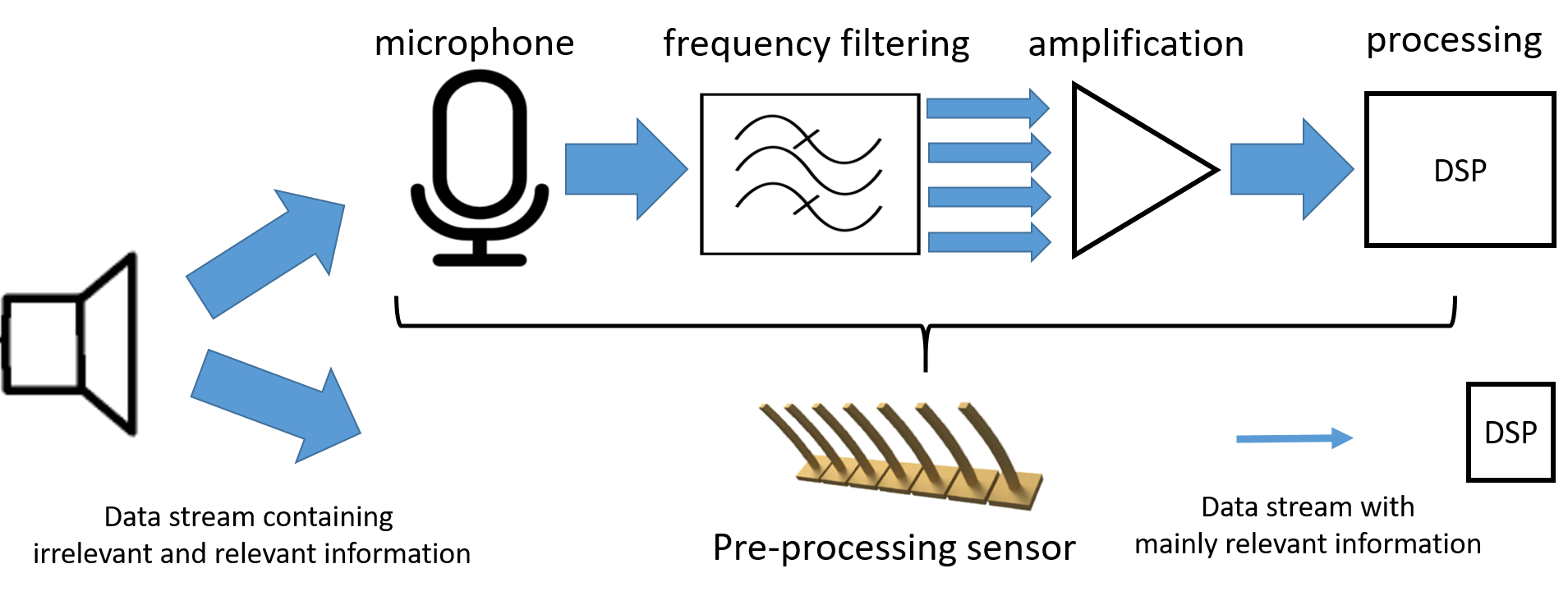}
\caption{Concept of pre-processing sound sensor}
\label{fig:concept}
\end{figure}

The nonlinear amplification can be tuned by modifying the dynamics with feedback loops. This enables (i) a pre-processing at the sensor level and (ii) the possibility to sense only the relevant sound sources by suppressing detection of irrelevant signals. The latter is particularly interesting considering  sensor data streamed to and analyzed by remote processing units. The amount of data streamed, particularly real-time data, is strongly increasing \cite{Data} due to concepts such as smart factories and Internet of Things. Adaptive and intelligent sensors, which only acquire the relevant information by suppressing noise and irrelevant data, could be an answer to this data handling problem (see Fig.~\ref{fig:concept}).

We aim at building nonlinear, adaptive, acoustic sensors that by design include a pre-processing of data and a tunable, nonlinear amplification similar to the biological ideal. Therefore, we use mechanical resonators with an integrated deflection sensing and actuation mechanism. Their dynamics shall be shifted into a nonlinear regime using a feedback loop, as was proposed by Joyce et al. \cite{Joyce}, or an output coupling loop. The loop feeds back either the sensor's output signal or the output signal of another sensor to the actuator of the considered sensor. The latter case is particularly interesting, since theoretical studies by Gomez et al. \cite{Gomez} show that coupling of sub-threshold Hopf oscillators can result in a stronger nonlinearity and a sharper response. Thus, the amplification of low sound pressure levels should be increased as well, enlarging the dynamic range of the sensors. In both cases, feedback or coupling, a transition from a resting state to an autonomous oscillatory state (without external forcing) should be observed in dependence of the feedback or coupling parameter \cite{Joyce, Gomez}. Near this transition the sound detection should become nonlinear \cite{Joyce}. We demonstrate the realization of such adaptive acoustic sensors, which could act as artificial hair cells with similar sensing properties as the biological ideal.  

\section{Sensor Design and Experimental Methods}
\begin{figure}[!t]
\centering
\includegraphics[width=3.5in]{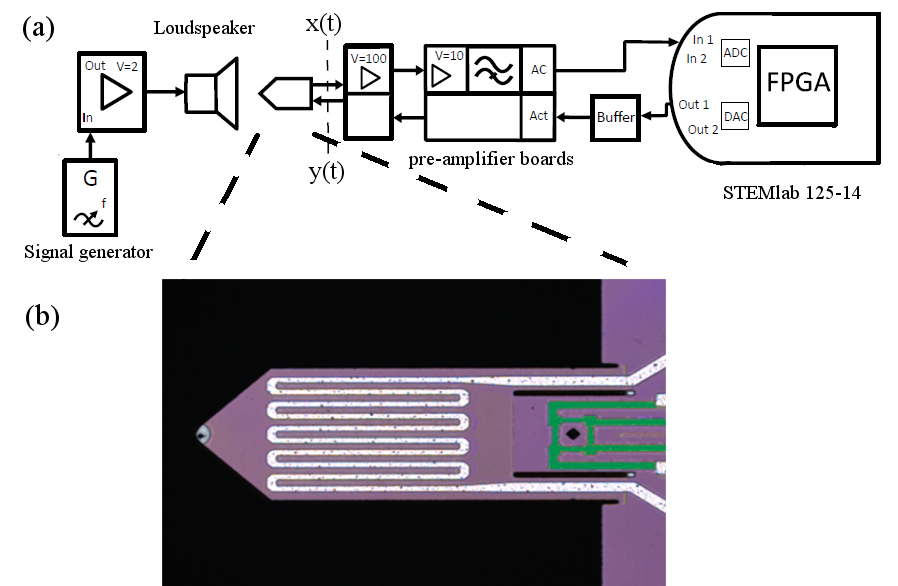}
\caption{(a) Schematic representation of system for sound detection with a nonlinear sensor, (b) Active cantilever used as sound sensors with thermo-mechanic actuation (white) and piezo-resistive deflection sensing (green).}
\label{fig:sensor}
\end{figure}

A sketch of the implemented system is shown in Fig.~\ref{fig:sensor}. The system consists mainly of two parts: first, the mechanical resonator with integrated actuation and deflection sensing and, second, the feedback loop for tuning the resonator dynamics. As mechanical resonator active cantilevers are used [see Fig.~\ref{fig:sensor}(b)], which are typically employed for atomic force microscopy \cite{Rangelow}. These are $350~\mathrm{\mu m}$ long, $150~\mathrm{\mu m}$ wide, and $2-4~\mathrm{\mu m}$ thick silicon beams. The top layer of this composite structure is an aluminum loop, which is used as thermo-mechanical actuator [marked in white in Fig.~\ref{fig:sensor}(b)]. Thereby, a current through the actuator results in a heating of the beam. This yields a deflection of the beam due to the different thermal expansion coefficient of silicon and aluminum. The deflection is proportional to the introduced power. Thus, a sine wave actuation signal yields an oscillation with twice the actuation frequency. For detection of the beam's deflection, piezo-resistive elements are embedded in a Wheatstone Bridge configuration near the base of the beam [marked in green in Fig.~\ref{fig:sensor}(b)]. A deflection of the beam changes the resistivity of these elements, resulting in a change of the readout voltage. The integration of actuation and sensing into the beam enables a miniature system, which can be immediately used without alignment and adjustment of readout.

The feedback loop, used to tune the dynamics of the beam, consists of (i) reading the sensor signal, (ii) calculating the feedback function and (iii) using the feedback signal to drive the actuator [see fig.~\ref{fig:sensor}(a)]. Here we use two different strategies for determining the feedback signal $y$. In the first case, a self-feedback mechanism is installed: the sensor signal $x$ is multiplied by the self-feedback parameter $a$ to obtain the feedback signal 
\begin{equation}
y(t)=a x(t),
\label{eq:feedback}
\end{equation} where $a>0$. This signal $y(t)$ is used to drive the actuator.

In the second case, a coupling of two sensors is realized by feeding the output signal of one sensor $x_1(t)$, multiplied by the coupling parameter $b$, as driving signal to the actuator of a second sensor and vice versa, i.e. 
\begin{equation}
y_2(t)=b x_1(t),\; \\
y_1(t)=b x_2(t).
\label{eq:coupling}
\end{equation}
Here, the indices $1$ and $2$ denote the respective sensor.

The calculation of the feedback signal is done in an FPGA architecture on a STEMlab 125-14 board, which allows a near real-time feedback (approx. $0.1~\mathrm{\mu s} - 1~\mathrm{\mu s}$ delay, corresponding to max. 1.4~\% of the oscillation period of the resonator). Before the feedback is calculated, the sensor signal is amplified by a factor of $1000$, high-pass filtered ($f_g\approx 1~\mathrm{kHz}$) to use only the AC signal, and finally digitized by an analog-to-digital converter on the STEMlab-board (sample rate 125~MHz and 14~bit resolution). While the input range of the STEMlab board can be switched between $\pm 1~\mathrm{V}$ and $\pm 20~\mathrm{V}$, the output range is limited to $\pm 1~\mathrm{V}$. Values of the feedback signal, outside of this range are mapped to the maximal value. After the calculation, the feedback signal is converted into an analog voltage signal by the digital-to-analog converter of the STEMlab board (sample rate 125~MHz), fed to a buffer board and finally applied to the actuator of the beam. The sensor signal is furthermore saved into a file (sample rate $1.98~\mathrm{MHz}$ or $0.49~\mathrm{MHz}$) and subsequently analyzed.

To study the dynamics of the sensor and the nonlinear amplification, first, the dynamics of the beam are studied in dependence of the self-feedback parameter $a$ and the coupling parameter $b$. Therefore, no sound is applied, while for the feedback signal the parameters $a,b$ are varied in a range of $0$ to $72$. For this case, the maximal peak-to-peak voltage is determined from the sensor signal. In a second step, the acoustic sensing properties are tested. Therefore, a piezo-loudspeaker (KEMO L010) is driven by a sine wave signal, which is generated by a signal generator (Agilent 33521 A) and pre-amplified by a factor of two. The signal contains a linear frequency sweep from $12.5~\mathrm{kHz} - 16~\mathrm{kHz}$ and back (sweep time $1.5~\mathrm{s}$, return time $1.5~\mathrm{s}$) with various driving voltage amplitudes $V_{\mathrm{sound}}$. The frequency response curve of the loudspeaker is approximately flat in this range.  

\section{Results}
\begin{figure}[!t]
\centering
\includegraphics[width=3.9in]{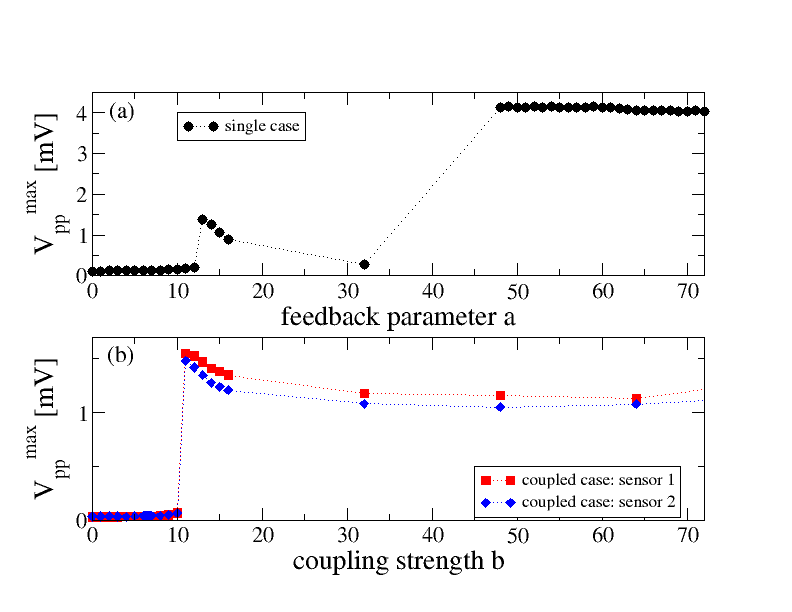}
\caption{Dependence of peak-to-peak amplitude of the sensor signal on (a) feedback parameter $a$ and (b) coupling parameter $b$.}
\label{fig:feedback-dep-a}
\end{figure}
First we will discuss the results describing the dynamics of the beam in dependence of the feedback and coupling strength without acoustic excitation. In Fig.~\ref{fig:feedback-dep-a}(a), the maximal peak-to-peak amplitude $V_{pp}^{max}$ of the sensor signal (black circles) is shown in dependence of self-feedback parameter $a$ (cf. eq.~\ref{eq:feedback}) for the case of a single sensor. In Fig.~\ref{fig:feedback-dep-a}(b), $V_{pp}^{max}$ is shown in dependence of the coupling parameter $b$ (cf. eq.~\ref{eq:coupling}) for the case of two coupled sensors. 

For the self-feedback case, two transition points are observed, $a \approx 12$ and $a \approx 46$, at which the amplitude increases strongly. In the regions between these points the amplitude is decreasing. Below $a\approx 12$, the mechanical resonator is in a resting state and the signal consists only of broadband noise. For $a>11$ an oscillating signal is measured with frequency contributions in the range of the first oscillation mode of the beam ($\approx 14~\mathrm{kHz}$) and a broader peak near $600~\mathrm{kHz}$. Thus, at $a=12$ a transition form quiescence to autonomous oscillation is observed. The observed behavior is an indication of a Hopf bifurcation, which is discussed as origin for the nonlinear amplification in human hearing \cite{Juelicher}. However, alternative mechanisms are also conceivable that can lead to the observed autonomous oscillation of the cantilever.

A similar behavior can be observed if two beams are coupled by their output signals, shown in Fig.~\ref{fig:feedback-dep-a}. Thereby, the frequencies of the first mode for the two beams are $13.7$~kHz and $14.2$~kHz with a quality factor $Q \approx 30$. As can be seen in Fig.~\ref{fig:feedback-dep-a}, a strong increase of the amplitudes of the sensors occurs for $b \approx 10$. The signals of both sensors exhibit oscillations for $b>10$, while for smaller $b<10$ the signals consist only of broadband noise. Note, the value of $b=10$ lies between the values of the self-feedback parameter $a$ ($a=8$ for sensor~1 and $a=12$ for sensor~2), at which the transition to autonomous oscillation mode occurs for each of the beams in the single beam, self-feedback case. Thus, self-feedback as well as coupling can be used to obtain a transition from quiescent to autonomous oscillations. 

\begin{figure}[!t]
\centering
\includegraphics[width=3.5in]{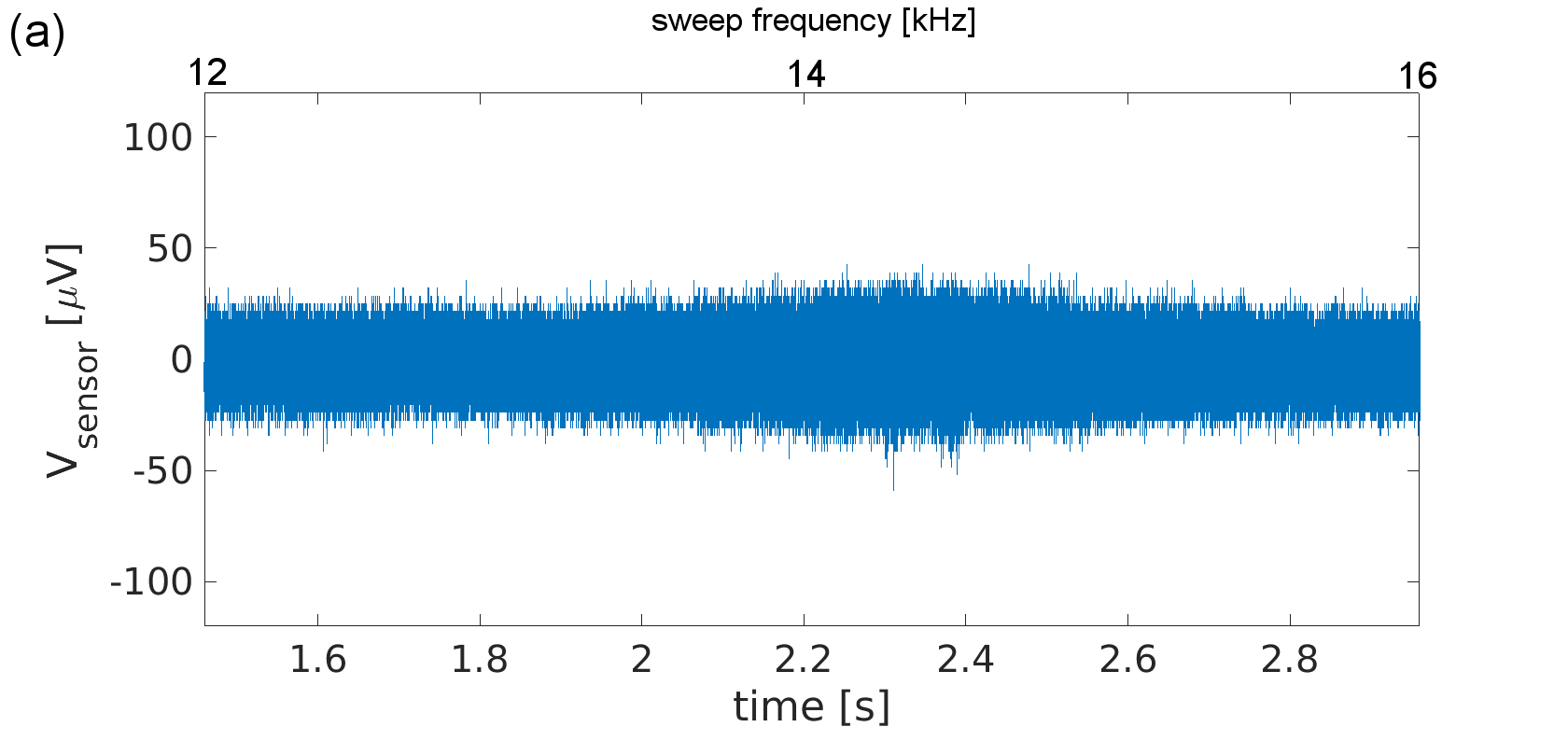}
 \includegraphics[width=3.5in]{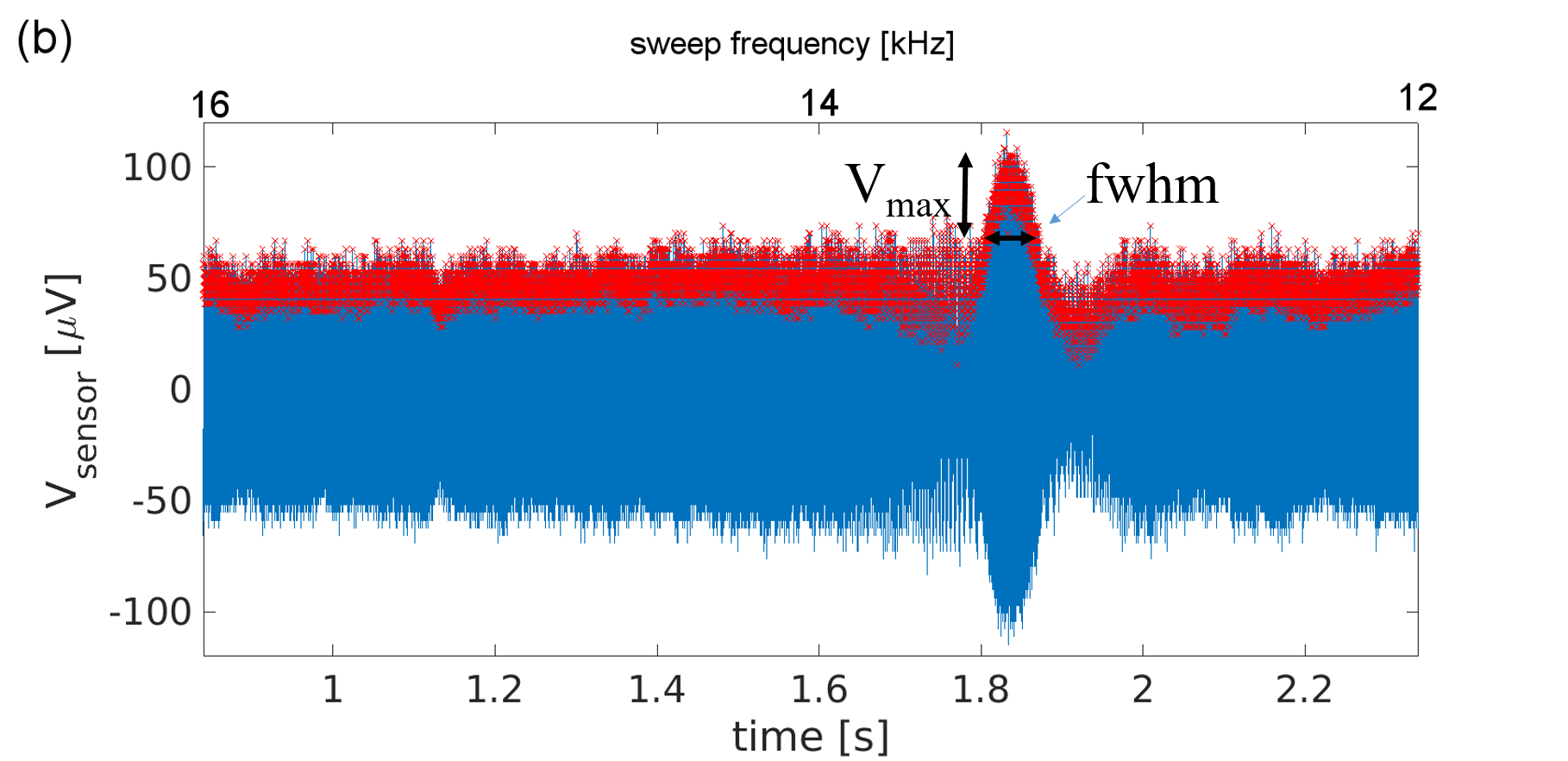}
\caption{Times series of sensor signal (blue) in response to a frequency-sweeped sound signal with $V_{\mathrm{sound}}=500$~mV for two different values of the self-feedback parameter $a$: (a) $a=0$ and (b) $a=63$. For better visualisation, the maximal values of the sensor signal (upper part of signals envelope) are marked in red.}
\label{fig:soundresp1}
\end{figure}
As described earlier, it was argued that the nonlinear dynamics of the hair cells resemble the dynamics of a critical oscillator tuned at the point of a Hopf bifurcation \cite{Juelicher}. In this nonlinear regime, nonlinear amplification of sound is automatically realized \cite{Juelicher}. Thus, the dynamics of the mechanical beams shall be tuned into this regime, i.e. $a\approx 12$ and/or $b\approx 10$. In Fig.~\ref{fig:soundresp1}, the times series of the sensor signal is shown for excitation with a sound signal for two different values of the self-feedback parameter $a$: (a) $a=0$ and (b) $a=63$. Here, a voltage of $500$~mV was used to drive the loudspeaker and the frequency of sound signal was increased linearly from $12.5$~kHz to $16.5$~kHz in $1.5$~s. The resonance of the sensor to a certain sound frequency can be observed in both cases by a maximum in signal amplitude. However, for the passive case [$a=0$ in Fig.~\ref{fig:soundresp1}(a)], this resonance amplitude is only slightly larger than the noise level. In contrast, for the self-feedback case a strong increase of the resonance amplitude is observed. This increase is much stronger than the increase of the noise level, resulting in a much larger signal-to-noise ratio than in the passive case. Besides the increase in resonance amplitude, the width of the peak decreases, too, which corresponds to an increased frequency resolution of the sensor. Thus, the self-feedback improves the acoustic sensing properties.

To quantify the change in sensing properties, the maximal resonance amplitude $V_{\mathrm{max}}$ and the resonance frequency are determined from the time series of the sensor signal. Thereby, $V_{\mathrm{max}}$ is the difference between the maximal noise level and the maximal resonance amplitude [see fig.~\ref{fig:soundresp1}(b)]. The resonance frequency is the frequency, at which the maximal sensor amplitude is measured. 
\begin{figure}[!t]
\centering
\includegraphics[width=3.9in]{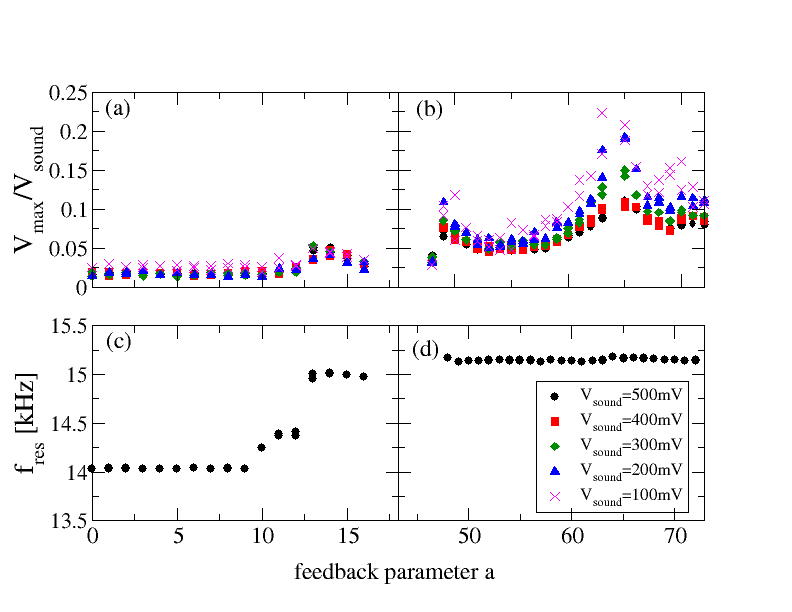}
\caption{Dependence of (a) ratio of maximal resonance amplitude $V_{\mathrm{max}}$ to driving voltage of loudspeaker $V_{\mathrm{sound}}$ and (b) resonance frequency $f_{\mathrm{res}}$ on the self-feedback parameter $a$ for various sound pressure levels and two different ranges of $a$: (a) and (c) $0<a<18$; (b) and (d) $44<a<72$.}
\label{fig:soundresp2}
\end{figure}
Both values are shown in Fig.~\ref{fig:soundresp2} for two different ranges of the self-feedback parameter $a$: (a) and (c) $a=[0,17]$, and (b) and (d) $a=[45,72]$, corresponding to the two transition regions. Different voltages $V_{\mathrm{sound}}$ were applied to drive the loudspeaker, corresponding to different sound pressure levels. The maximal resonance amplitude was divided by the driving voltage of the loudspeaker to analyse the dependence of sensing on the sound pressure level. For $a<12$, $V_{\mathrm{max}}/V_{\mathrm{sound}}$ is quite small and almost independent of the sound pressure level, as it is expected for passive resonance. For $a>12$ the response of the sensor is slightly increasing near $a=12$ but strongly increasing for $a\approx 63$. Thereby, the response of the sensor is strongly depending on the sound pressure level for $a\approx 63$ (but not for $a\approx 12$). In particular, the strongest relative amplification is observed for the signal with lowest sound pressure level, as can be seen from the value of $V_{\mathrm{max}}/V_{\mathrm{sound}}$. This is expected for the nonlinear amplification of an oscillator tuned near the Hopf bifurcation point. Furthermore, this nonlinear amplification results in a larger dynamic range, since for $a\approx 63$ the resonance of the sensor can be detected from its time series even for very low driving voltages  $V_{\mathrm{sound}}=100$~mV of the loudspeaker (i.e. sound pressure levels). In the passive case without feedback, the resonance is covered by the noise of the signal. Regarding the resonance frequency, one observes a shift of the resonance frequency from $14.2$~kHz for $a<10$ to $15.1$~kHz for $a>12$. 

\newpage
\section{Conclusions}
We have demonstrated that the dynamics of mechanical resonators can be tuned into a nonlinear dynamics regime using self-feedback or output signal coupling. Above the transition, nonlinear sensing of sound signals is observed. Thereby, the self-feedback yields \begin{itemize}                                                                                                                                                                                                                                              \item an increase of the resonance amplitude,                                                                                                                                                                                                                                                                 \item an increase of the signal-to-noise ratio,                                                                                                                                                                                                                                                                 \item a decrease of the width of the resonance peak, corresponding to an improved frequency resolution,                                                                                                                                                                                                                                                                 \item an increase of the dynamic range enlarging the range to lower sound pressure levels and                                                                                                                                                                                                                                                        \item a shift of the resonance frequency.                                                                                                                                                                                                                                  \end{itemize}
                                                                                                                                                                                                                                                                
With such properties, a silicon-based micro-mechanical resonator is a promising solution for developing artificial hair cells. Thereby, changing the self-feedback parameter~$a$ enables tuning the sensing properties, in particular the amplification and sensitivity. Combining the tunable acoustic, nonlinear sensors with a sound analysis, whose outcome is used to tune the sensor properties, would be a route for adaptive sensors, which acquire only the relevant sound information.

\section*{Acknowledgment}
We gratefully acknowledge financial support by the Deutsche Forschungsgemeinschaft (DFG) in the frame of Research Unit RU 2093 'Memristive Devices for Neural Systems' and the Bundesministerium f\"ur Bildung und Forschung (BMBF) in the frame of 'Forschungslabor Mikroelektronik Deutschland: ForLab NSME' (FKZ: 16ES0939).



%

\end{document}